\documentclass[referee]{aa}

\def\gsim{ \lower .75ex \hbox{$\sim$} \llap{\raise .27ex \hbox{$>$}} }
\def\lsim{ \lower .75ex\hbox{$\sim$} \llap{\raise .27ex \hbox{$<$}} }
\def\observed{{\it observed}\ }

\usepackage{graphicx}
\authorrunning{Guetta \& Piran}

\titlerunning{Long duration GRBs}

\begin{document}
\title{Do long-duration GRBs follow star formation?}
\author{Dafne Guetta\inst{1}
\and Tsvi Piran\inst{2}}

\institute{ Osservatorio astronomico of Rome v. Frascati 33 00040
Monte Porzio Catone, Italy  \and  Racah Institute for Physics, The
Hebrew University, Jerusalem 91904, Israel}
\date{To be determined}

\abstract{ We compare the luminosity function and rate inferred
from the BATSE long  bursts peak flux distribution with those
inferred from the {\it Swift} peak flux distribution. We find that
both the BATSE and the {\it Swift} peak fluxes can be fitted by
the same luminosity function and the two samples are compatible
with a population that follows the star formation rate. The
estimated local long GRB rate (without beaming corrections) varies
by a factor of five from 0.05 Gpc$^{-3}$yr$^{-1}$ for a rate
function that has a large fraction of high redshift bursts to 0.27
Gpc$^{-3}$yr$^{-1}$ for a rate function that has many local ones.
We then turn to compare the BeppoSax/HETE2 and the {\it Swift}
observed redshift distributions and compare them with the
predictions of the luminosity function found. We find that the
discrepancy between the BeppoSax/HETE2 and {\it Swift} observed
redshift distributions is only partially explained by the
different thresholds of the detectors and it may indicate strong
selection effects. After trying different forms of the star
formation rate (SFR) we find that the observed {\it Swift}
redshift distribution, with more observed high redshift bursts
than expected, is inconsistent with a GRB rate that simply follows
current models for the SFR. We show that this can be explained by
GRB evolution beyond the SFR (more high redshift bursts).
Alternatively this can also arise if the luminosity function
evolves and earlier bursts were more luminous  or if strong
selection effects affect the redshift determination.

} \maketitle

\section{Introduction}

Gamma ray bursts (GRBs) are one of the most powerful events in the
universe. The high energy photons emitted travel from cosmological
distance tracing the star formation history in the universe. Our
understanding of long ($T_{90}>2$sec \footnote{$T_{90}$ is defined
as the time interval in which 90\% of the prompt energy arrives.})
GRBs and their association with stellar collapse follows from the
discovery in 1997 of GRB afterglow and the subsequent identification
of host galaxies, redshift measurements and detection of associated
Supernovae.

However the number of GRBs with a measured redshift is still
limited. Only a small fraction  of the BeppoSax/HETE2 bursts have
measured redshifts. It was expected that {\it Swift} would allow
further insight into the redshift properties of these objects.
Indeed, the ability of {\it Swift} to locate and follow-up fainter
bursts than the previous satellites, has allowed more distant bursts
to be studied.  The mean redshift of the BeppoSax/HETE2 sample was
$z_{mean}=1.4$, while bursts discovered by {\it Swift} now have
$z_{mean}=2.8$. The number of {\it Swift} bursts with a measured
redshift is still small as only $\sim 30$ bursts out of 130 detected
bursts have a known redshift. The selection effects that arise in
both samples are not clear and hard to quantify (Fiore et al. 2006).
Therefore, at present we cannot derive directly the GRB luminosity
function and rate evolution that are fundamental to understand the
nature of these objects.

We can constrain the luminosity function and rate distribution by
fitting the  BATSE and {\it Swift} peak flux distributions  to those
expected for a given luminosity function and GRB rate (Piran 1992,
Cohen \& Piran 1995, Fenimore \& Bloom 1995, Loredo \& Wasserman
1995, Horack \& Hakkila 1997, Loredo \& Wasserman 1998, Piran 1999,
Schmidt 1999, Schmidt 2001, Sethi \& Bhargavi 2001, Guetta et al.
2005, Guetta \& Piran 2005, 2006). Since the observed flux
distribution is a convolution of these two unknown functions we must
assume one and find a best fit for the other. We assume that the
rate of long bursts follows the star formation rate (or a
modification of the star formation rate, discussed later) and we
search for the parameters of the luminosity function. We show, in
the first part of the paper that one can obtain a fully consistent
fit for both the BATSE and the {\it Swift} peak flux samples.

A more complicated issue is to compare the observed redshift
distribution of BeppoSax/HETE2 bursts with  the observed redshift
distribution of {\it Swift} bursts and with the predictions of the
models for the rate and luminosity function that were inferred from
the peak flux distributions. We turn to this problem in the second
part of the paper.  Clearly the {\it intrinsic} GRB distribution is
the same and the differences between the observed distributions
should arise from the differences in thresholds, in the observed
energy band and from selection effects. These factors determine
together  the samples of bursts with observed redshifts. Among these
factors the issue of selection effects is least understood  (see
however, Hogg and Fruchter, 1999; Bloom et al. 2003 and Fiore et
al. 2007). We attempt to correct the
BeppoSAX/HETE2 sample for some of the known selection effects.  We
also consider two limiting cases in which all the bursts with
missing redshifts are either at very low or very high redshifts. 

  Due to  the higher detection thresholds the
BeppoSax/HETE2 {\it observed} distribution is nearer to us than the
{\it Swift} one. However, the difference in thresholds is not enough
to explain the difference between the two observed redshift
distributions. With more {\it Swift}  high redshift bursts than
expected we conclude that either GRBs evolve faster than the SFR
(more high redshift bursts), or that the assumption that the
luminosity function is independent of z is wrong.  Daigne et al.
(2005) consider these possibilities and, following a different
approach, obtain similar results. We also consider models where the
GRB rate is a convolution of the SFR and of a sharp jump  in the
rate to high values at high redshift. These models could be related
to the fact that GRBs seem to be more abundant  in low metallicity
regions (Fynbo et al., 2003; Vreeswijk et al., 2004). Such a jump
could arise, for example, from a low metallicity threshold, below
which the GRB rate jumps.   This possibility has been explored
by Natarajan et al. (2005) whose results are similar to the ones
obtained in this work. Nuza et al. (2007) developed a Monte Carlo
code that simulates a long GRBs distribution for a model where only
low-metallicity massive stars are long GRBs progenitors. The results
of their calculations are also in agreement with the ones found in
our work. Note, however, that a recent analysis of Fynbo et al.
(2006) shows that GRBs occur in environments covering a broad range
of metallicity at a given redshift. Alternatively it is
possible, but less likely, that the observed distribution is
completely determined by still unknown selection effects which
dominate the BeppoSAX/HETE2 and the {\it Swift} redshift
determination in unexpected manners.

\section{Luminosity function  from the BATSE and the {\it Swift} samples}

Our  methodology follows Guetta et al. (2005). For BATSE we consider
all the long GRBs detected while the BATSE onboard trigger was set
for 5.5 $\sigma$ over background in at least two detectors, in the
energy range 50-300 keV. Among those we took the bursts for which
$C_{\rm max}/C_{\rm min} \geq 1$ at the 1024 ms timescale, where
 $C_{\rm max}$ is the count rate in the second brightest illuminated detector
and $C_{\rm min}$ is the minimum detectable rate. These constitute a
group of 595 bursts. In our previous paper (Guetta et al. 2005) we
have shown that the distribution of minimal rates is very narrow and
we can take an average rate that corresponds to a threshold
$P^{(50-300)keV}_{\rm lim, BATSE}\sim$ 0.25 ph cm$^{-2}$ s$^{-1}$.

For {\it Swift} we consider all long bursts detected until September
2006 ($\sim 130$) bursts in the energy range 15-150 keV. {\it
Swift}'s complicated triggering algorithm is not based just on the
concept of a minimal flux above the background. Still we can have an
estimate of the effective {\it Swift} threshold by plotting in Fig.
1 the peak flux distribution. This figure compares the peak-flux
cumulative distributions of the {\it Swift} GRBs with that of BATSE.
The comparison is done in the energy band 50-300 keV, which is the
band where BATSE detects GRBs. Note that for this comparison we have
converted the BAT 15-150 keV peak fluxes to fluxes in the 50-300 keV
band using the BAT peak fluxes and spectral
parameters\footnote{These parameters were taken from the {\it Swift}
information page
http://swift.gsfc.nasa.gov/docs/swift/archive/grb\_table.html}. We
find an effective threshold of $P^{(50-300)keV}_{\rm lim, Swift}\sim
0.18$ ph cm$^{-2}$ s$^{-1}$ in agreement with what found by
Gorosabel et al. (2004).  Note that within the {\it Swift} band this
threshold corresponds to $P^{(15-150)keV}_{\rm lim, Swift}\sim 0.4$
ph cm$^{-2}$ s$^{-1}$. The sensitivity of BeppoSax, HETE2 and {\it
Swift} has been studied in detail by Band (2003, 2006). Band (2006)
also studied the sensitivity of BAT instrument as a function of the
combined GRB temporal and spectral properties. We refer to these
papers for more details on these topics. The values of sensitivities
used in this paper are in agreement with Band's results.

\begin{figure}[h]
{\par\centering \resizebox*{0.85\columnwidth}{!}{\includegraphics
{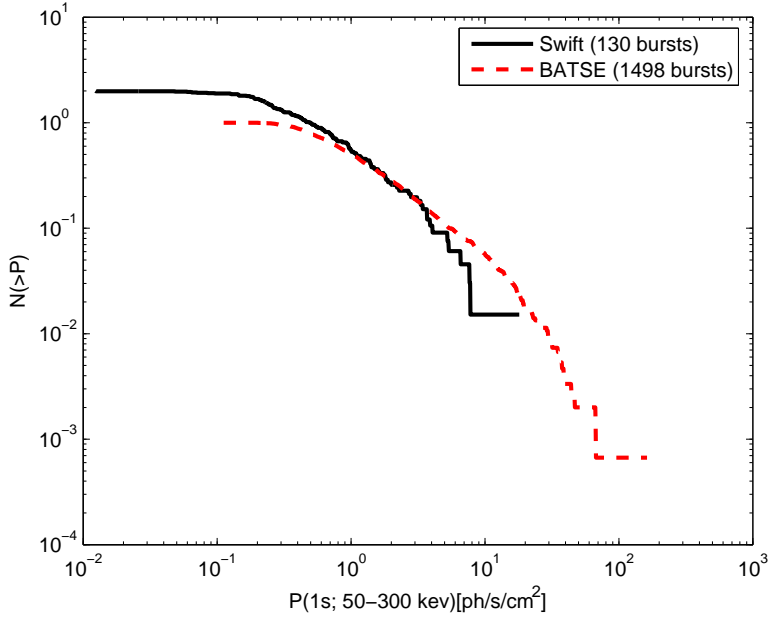}} \par} \caption{\label{fig1} The observed
BATSE and {\it Swift} peak flux cumulative  distributions in the 50-300 keV
energy range}
\end{figure}


\begin{figure}[h]
\begin{tabular}{cc}
\includegraphics[height=6.8truecm,width=6.8truecm]{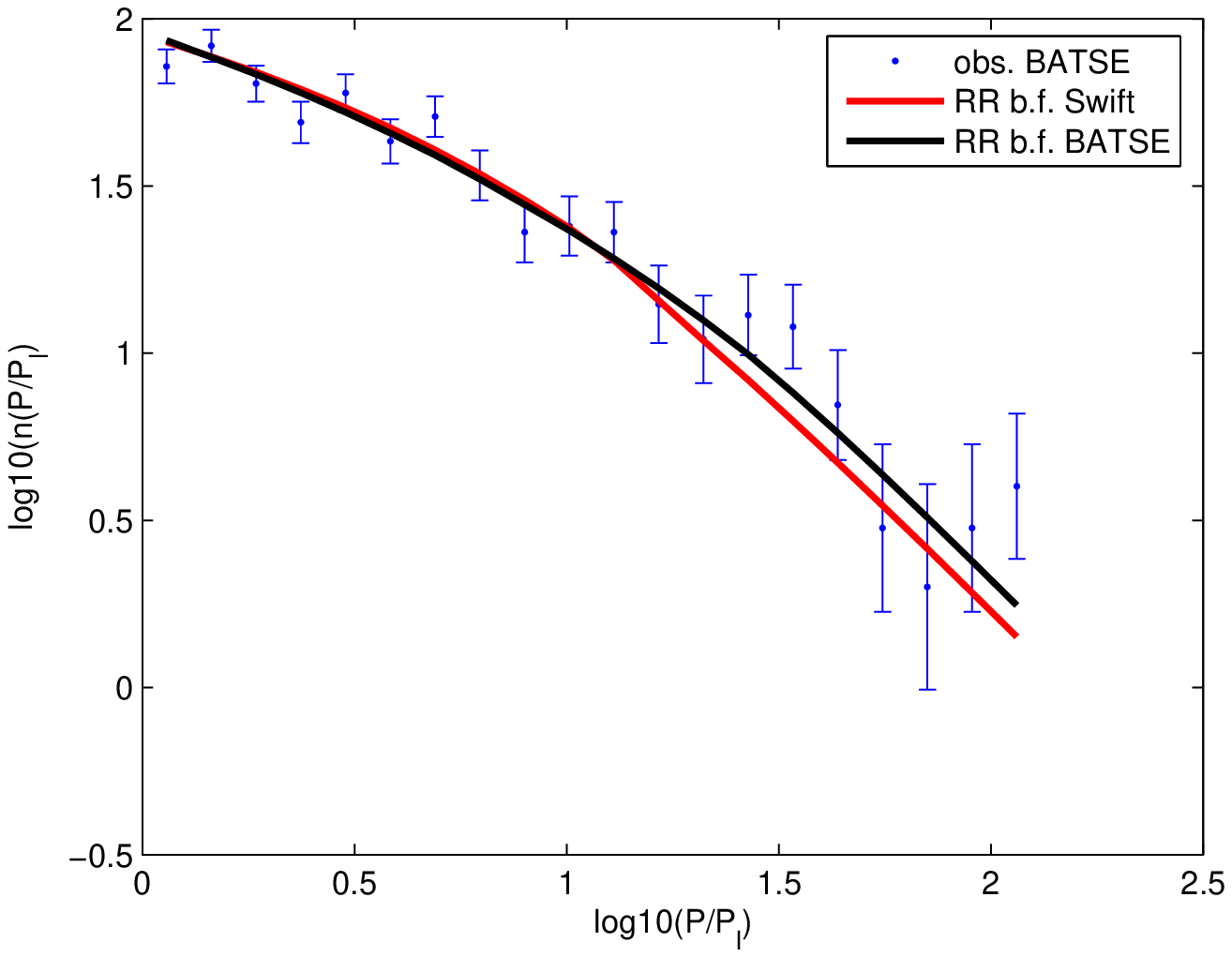}
\includegraphics[height=6.8truecm,width=6.8truecm]{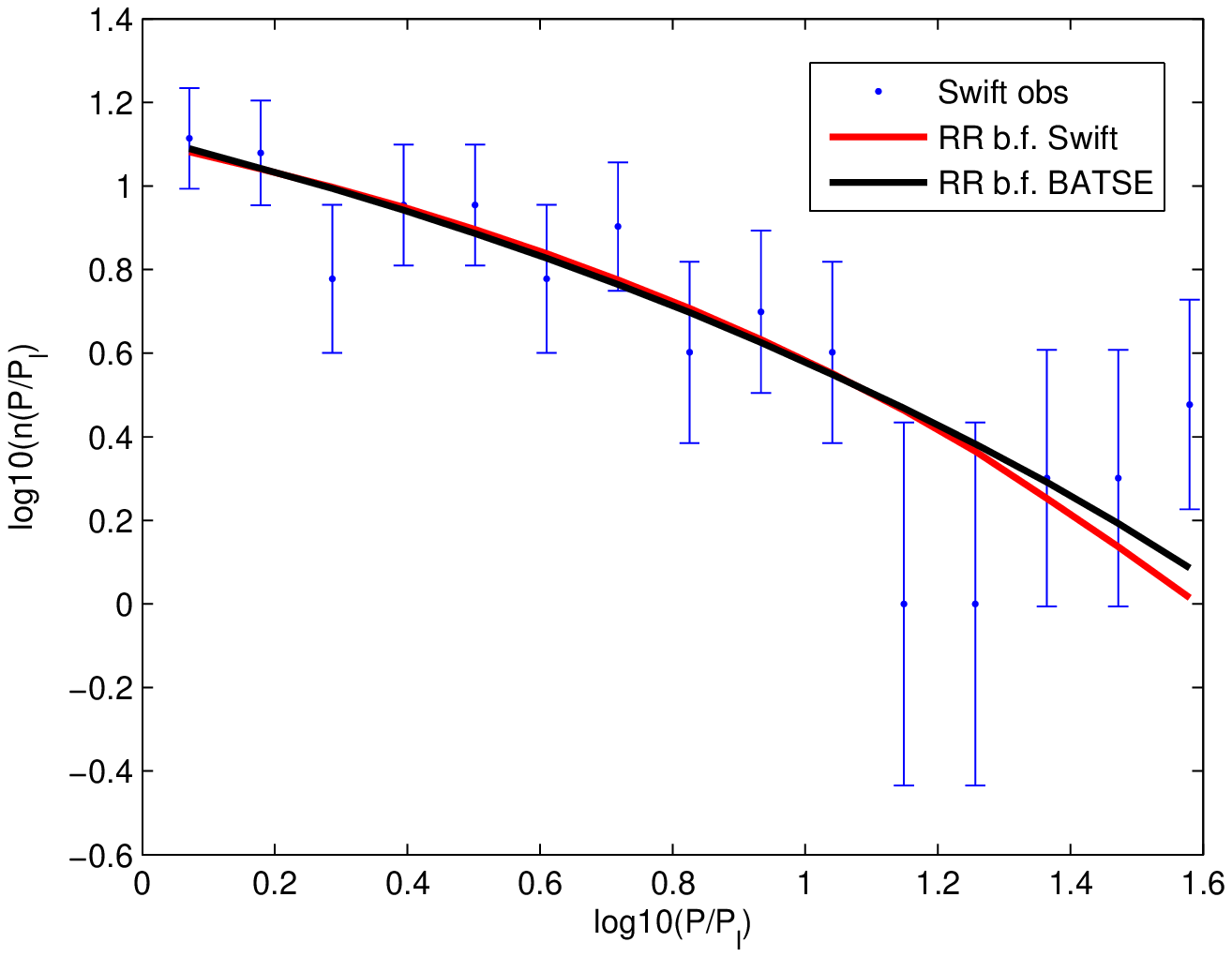}
\end{tabular}
\caption{\label{diffBS} a) left panel: The predicted differential
distribution (n(P/P$_{\rm lim}$)  with the luminosity function
parametrs that best fit the BATSE sample (black curve) and the
{\it Swift} sample (red curve)  with a RR-SFR vs. the  observed
n(C$_{\rm max}$/C$_{\rm min}$) taken from the BATSE catalog. b)
right panel: The predicted differential distribution (n(P/P$_{\rm
lim}$)  with the luminosity function parametrs that best fit the
BATSE sample (black curve) and the {\it Swift} sample (red curve)
with a RR-SFR vs. the  observed {\it Swift} n(P/P$_{\rm lim}$)
taken from the {\it Swift} catalog}
\end{figure}

The method used to derive the luminosity function is essentially
the one used by Schmidt (1999) and by Guetta et al. (2005). We
consider a broken power law with lower and  upper limits which are
factors of $1/\Delta_1$ and $\Delta_2$ respectively times the
break luminosity $L^*$. The luminosity function (of the  peak
luminosity $L$) in the interval $\log L$ to $\log L + d\log L$ is:
\begin{equation}
\label{Lfun}
\Phi_o(L)=c_o
\left\{ \begin{array}{ll}
(L/L^*)^{\alpha} &  L^*/\Delta_1 < L < L^* \\
(L/L^*)^{\beta} & L^* < L < \Delta_2 L^*
\end{array}
\right. \;,
\end{equation}
where $c_o$ is a normalization constant so that the integral over
the luminosity function equals unity. We stress that the
luminosity considered here is the ``isotropic" equivalent
luminosity, which is the one relevant for detection. It does not
include a correction factor due to beaming.

Assuming that long GRBs follow the star formation rate we employ
four parametrization of the star formation
rate:\hfill\break (i) Model SF2 of Porciani \& Madau (2001):
\begin{equation}
\label{SFR} R_{GRB}(z)=R_{\rm SF2}(z)=  \rho_0 \frac{23
\exp(3.4z)}{\exp(3.4z)+22}
\end{equation}
where $\rho_0$ is the GRB rate at $z=0$. \hfill\break  (ii) The
Rowan-Robinson SFR (Rowan-Robinson 1999: RR-SFR) that can be
fitted with the expression:
\begin{equation}
\label{RR}
R_{GRB}(z) = \rho_0
\left\{ \begin{array}{ll}
10^{0.75 z} & z<1 \nonumber \\
10^{0.75 z_{\rm peak}} & z\geq 1.
\end{array}
\right.
\end{equation}
 \hfill\break (iii) Model SF3 of Porciani \& Madau (2001)
with more star formation at early epochs:
\begin{equation}
\label{SFR} R_{GRB}(z)=R_{\rm SF3}(z)=  \rho_0 \frac{16
\exp(3.4z-0.4)}{\exp(-0.4)(\exp(2.93z)+15)}
\end{equation}
\hfill\break (iv)   The Star formation history parametric fit to
the form of Cole et al. (2001) taken from Hopkins and Beacon
(2006). This  rate strongly declines with z at high redshift
($z>4$)
\hfill\break  (v) We consider a toy model where the rate is
enhanced at high redshift and the transition is sharp. As
mentioned earlier this model might be related, to the lower
metallicity at higher redshift.  As a toy model we used a SFR as
described in model (ii) but at $z>2.5$ is enhanced by a factor of
2:
\begin{equation}
\label{RRmet} R_{GRB}(z) = \ \left\{ \begin{array}{ll}
R_{\rm RR}(z) & z<2.5 \nonumber \\
3\, R_{\rm RR}(z) & z\geq 2.5.
\end{array}
\right.
\end{equation}
Note that this model resembles but is not the same as the one used
by Natarajan et al. (2006) where there is just a jump in the rate
and no association with the SFR at all (as not enough details
about the model were given in that paper we could not reproduce it
here).

\begin{figure}[h]
{\par\centering \resizebox*{0.85\columnwidth}{!}{\includegraphics
{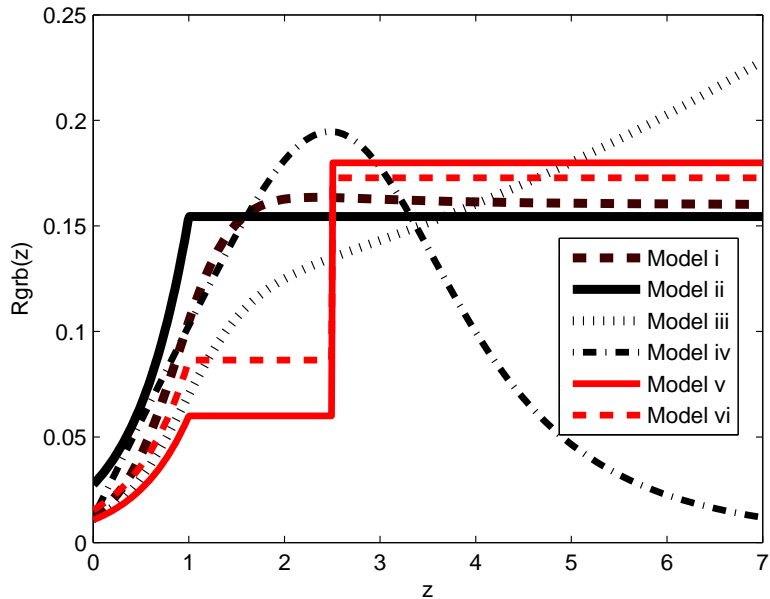}} \par} \caption{\label{rates} The star formation
histories considered in the paper }
\end{figure}

\hfill\break  (vi) For completeness we consider also a GRB rate that
is in between of model (ii) and (v)
\begin{equation}
\label{RRmet} R_{GRB}(z) = \ \left\{ \begin{array}{ll}
R_{\rm RR}(z) & z<2.5 \nonumber \\
2\, R_{\rm SF2}(z) & z\geq 2.5.
\end{array}
\right.
\end{equation}
We use the standard cosmological parameters $H_0 = 65~$ km s$^{-1}$
Mpc$^{-1}$, $\Omega_M = 0.3$, and $\Omega_{\Lambda} = 0.7$. The
different rates are shown in Fig \ref{rates}.

An important factor  is the cosmological k correction. We
approximate the typical effective spectral index in the observed
range of 50 keV to 300 keV as ($N(E)\propto E^{-1.6}$). This value
that was calculated by Schmidt (2001) for BATSE bursts is also
adequate for {\it Swift} bursts whose average spectral index is
$\sim 1.5$. The use of this average correction is justified when we
compare estimates of the luminosity based on this average value and
on the real spectrum (see Fig. \ref{zdistBATSE} below.)

\begin{figure}[h]
\begin{tabular}{cc}
\includegraphics[height=6.8truecm,width=6.8truecm]{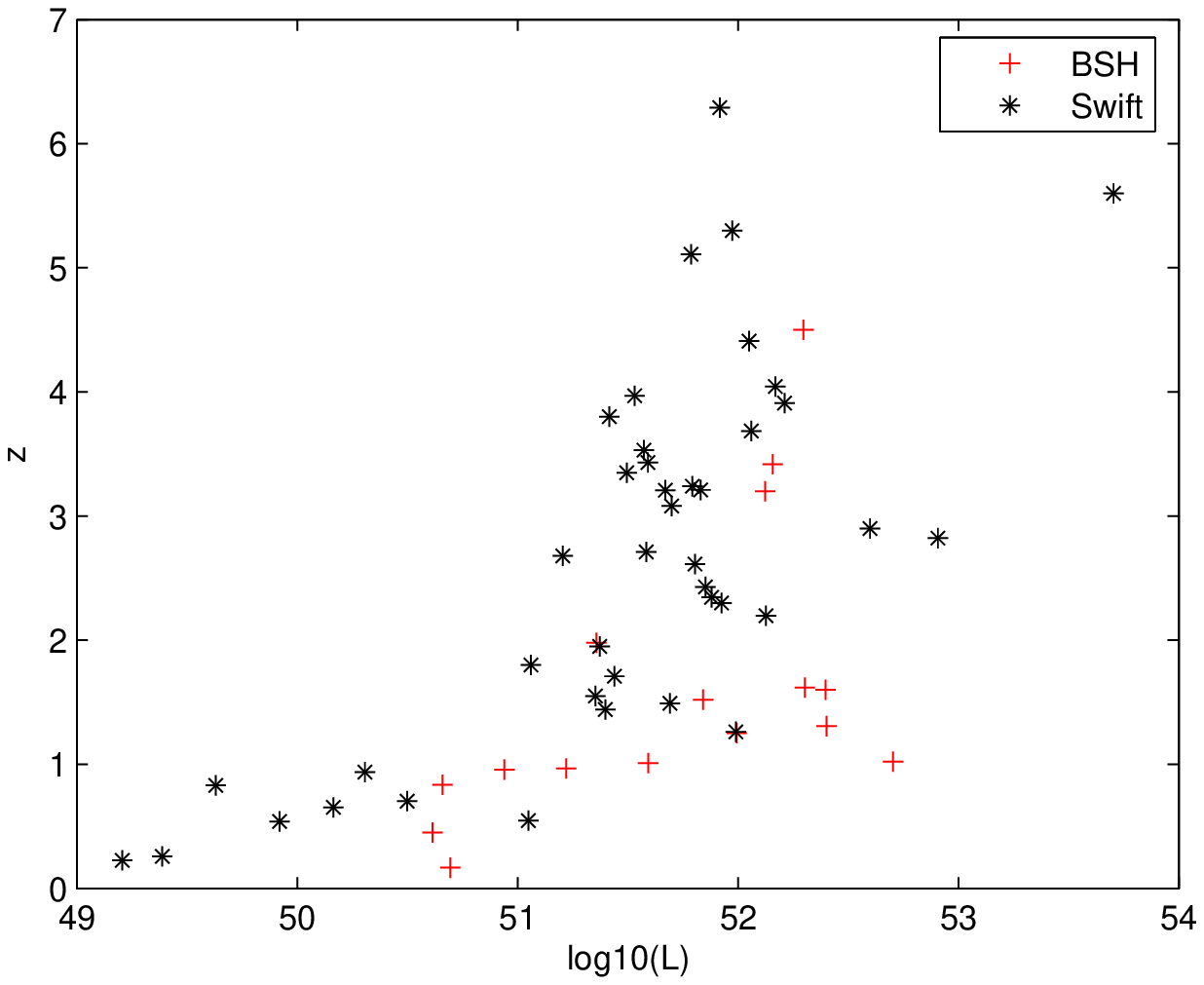}
\includegraphics[height=6.8truecm,width=6.8truecm]{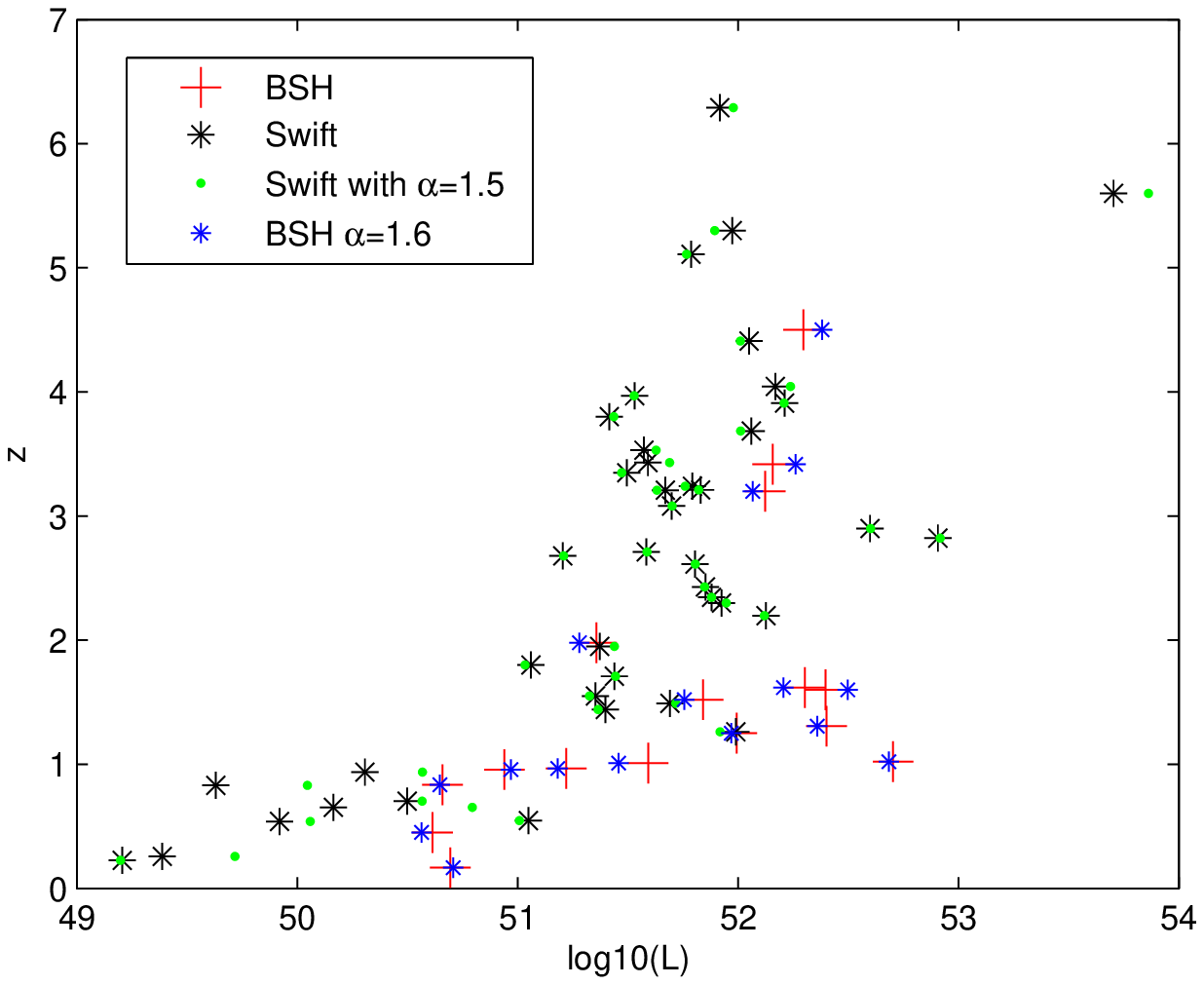}
\end{tabular}
\caption{a), left panel: the luminosities and redshifts of the
BeppoSAX/HETE2 (BSH in the legend) sample compared with the {\it
Swift} sample. b), right panel: the same as left panel, also marked
are the values of L extrapolated using the average photon index for
each sample.} \label{zdistBATSE}
\end{figure}

 The number of bursts with a peak flux $>P$ is given by:
\begin{eqnarray}
\nonumber
  N(>P)=\int\Phi_o(L)d\log L \int_0^{z_{max}(L,P)} \frac{R_{GRB}(z)}{1+z}
\frac{dV(z)}{dz}dz \
\end{eqnarray}
where the factor  $(1+z)^{-1}$ accounts for the cosmological time
dilation and $dV(z)/dz$ is the comoving volume element.

We find the best-fitting LF parameters and their dispersion by
$\chi^2$ minimization. We  vary the luminosity function parameters
$\alpha$, $\beta$, and $L^*$ keeping $\Delta_1=100$ and
$\Delta_2=100$ and inspect the quality of the fit to the observed
BATSE peak flux distribution. Once we obtain the best fit parameters
for the BATSE sample we test the quality of the fit with both the
observed BATSE and the {\it Swift} peak flux distributions. We then
repeat the same procedure and look for the luminosity function
parameters that best fit the {\it Swift} sample. Then we check the
quality of the fit with these parameters with the observed BATSE and
the {\it Swift} peak flux distributions.  Fig.  2a(2b) depicts a
comparison of the observed differential distributions $n(P)\equiv
dN/dP$  of BATSE ({\it Swift})  with the predicted distribution
obtained with the RR-SFR (model (ii)) for the parameters that best
fit the BATSE({\it Swift}) data. Also shown on the same figure  is
the curve obtained using the parameters that best fit the {\it
Swift}(BATSE) data. Similar curves are obtained for the other
models. The $\chi$-square values reported in Table I show the
consistency between the {\it Swift} and BATSE peak flux samples. The
consistency is reassuring. However the fact that we obtain good fits
for the data with very different models for the GRB rates reflects
the insensitivity, noticed already by Cohen and Piran (1995),  of
the peak flux distribution to the details of the GRB rate. The peak
flux distribution is a convolution of the luminosity function and
the GRB rate and different assumptions on the rate simply result in
different luminosity function.

The results of the fit are reported in Table 1. These show that the
best fit parameters $\alpha$ and $\beta$ are rather robust and they
do not depend on the exact shape of the GRB rate chosen and the
values of $\alpha$ and $\beta$ found for these rather different
models are all within the error bars of each other. To obtain the
local rate of GRBs per unit volume, $\rho_0$ we need to estimate the
effective full-sky coverage of the GRB samples. For BATSE we use 595
(47\% of the long GRBs)  events  detected over 1386 days in the
50-300 keV channel with a sky exposure of 48\%. For {\it Swift} we
consider  $\sim 130$ bursts detected in 1.5 yr and a sky coverage is
1/6. The value of $L^*$ is around $2.5 - 5.5\times 10^{51}$erg/sec.
It is somewhat higher for models (v) and (vi), as expected because
in these cases the intrinsic distribution is farther, and hence
stronger pulses are needed. The local rate varies, correspondingly
by a factor of 5 from the models  (v) and (vi)  to model (ii).

The fraction of expected high redshift ($z>6$) {\it Swift} bursts
vary strongly among the different models: (i) 1.3\%, (ii) 0.67\%
(iii) 3.5\%, (iv) 0.07\%, (v) 6.2\% and  (vi) 6.0\%. These results
are expected in view of the nature of the intrinsic distributions
that we consider in these models. The effect of the artificial
enhancement of the rate of high redshift bursts in models (v) and
(vi) is clearly seen.

It is interesting to compare the fraction of high redshift bursts,
that we find,  with previous attempts to estimate this number. Using
the Amati-like relation $E_p\propto L^{0.43}$ Daigne et al. (2005)
find 2.5 \% for model (i) (SF2-sfr) and 15 \% for model (iii)
(SF3-sfr) that are somewhat higher than our results. The discrepancy
with Daigne et al., (2005) may reflect the inapplicability of the
Amati relation (see Nakar and Piran (2005)). The fraction of bursts
at $z>4$ of our model (v) is  40\%,
 similar to the result obtained by Natarajan et al. (2005)
for their model (iv). Natarajan et al (2005) do not specify the
parameters of their model however they also consider an enhancement
in the redshift rate at $z\sim 3$ and the fact that we find similar
results is reassuring.  Bromm and Loeb (2006) consider the
contribution  of Pop III to high redshift bursts and find that 10\%
of all Swift bursts can originate at $z>5$. This is equal to what we
obtain with our models (v) or (vi). However, it is not clear if this
comparison is not a mere coincidence as our models involve a high
redshift enhancement in a constant factor above $z>2.5$ and one
don't expect Pop III stars at such ``low" redshifts.

The somewhat arbitrary values of $\Delta_{1,2}=(100,100)$ are chosen
in such a way that even if there are bursts less luminous than
$L^*/\Delta_1$ or more luminous than $\Delta_2 L^*$ they will
constitute only a very small fraction (less than $\sim 1\% $) of the
\observed bursts. Bursts above $L^* \Delta_2$ are very bright and
are detected to very large distances. However, such strong bursts
are very rare. Increasing $\Delta_2$ does not add a significant
number of bursts (observed or not) and this does not change the
results. In particular it does not change the overall rate.
$\Delta_1$ is more subtle.  The luminosity function increases
rapidly with decreasing luminosity. Thus, a decrease in $\Delta_1$
has  a strong effect on the overall rate of  GRBs. However, most of
the bursts below $L^*/\Delta_1$ are undetectable by current
detectors, unless they are extremely nearby. Even if the luminosity
function continues all the way to zero, this will increase
enormously the over all rate of the bursts (Guetta \& Piran 2006;
Guetta \& Della Valle 2006) (which will in fact diverge in this
extreme example)  most of these additional weak bursts will be
undetected and the total number of detected bursts won't increase.

\begin{table}[h]
\begin{tabular}{|c|c|c|c|c|c|c|}
 \hline
model-sample
& Rate(z=0)  & $L^*$ & $\alpha$ & $\beta$ &$\chi^2_{b.f.}$
&$\chi^2_{\rm other}$\\
   & $Gpc^{-3} yr^{-1}$ & $10^{51}$ erg/sec &  & & &  \\
  \hline
(i)-BATSE& $0.07^{+0.1}_{-0.05}$ & $5.5^{+2.1}_{-3.1}$ & $0.3^{+0.3}_{-0.2}$
&$2^{+1}_{-0.5}$ & $0.82$ & 1.2 \\
(i)-{\it Swift}& $0.10^{0.08}_{-0.06}$ & $3.3_{-1.0}^{+3.1}$ &
$0.1_{-0.05}^{+0.5}$ &$2_{-0.4}^{+0.8}$ & $0.85$ & 1.0 \\
(ii)-BATSE & $0.18^{+0.21}_{-0.1}$ & $5.5^{+2.1}_{-3.7}$&$0.4^{+0.2}_{-0.3}$ &
$2.5^{+0.5}_{-1}$&$0.86 $ &1.1  \\
(ii)-{\it Swift} &  $0.27^{+0.15}_{-0.22}$ & $2.3^{+5.1}_{-0.3}$ &
$0.1^{+0.5}_{-0.05}$ & $2^{+0.4}_{-0.5}$ & $0.81$ & $0.97$\\
(iii)-{\it Swift} &  $0.1^{+0.05}_{-0.03}$ & $4^{+2}_{-1}$ &
$0.1^{+0.3}_{-0.03}$ & $2^{+1}_{-0.2}$ & $0.82$ & $1.1$\\
(iv)- {\it Swift} & $0.11^{+0.08}_{-0.04}$ & $3.0^{+1.9}_{-2.8}$
& $0.2^{+0.3}_{-0.1}$ & $2^{+0.7}_{-0.5}$ & 0.83 & 1.2 \\
(v)- {\it Swift} & $0.05^{+0.03}_{-0.03}$ & $6.5^{+0.8}_{-2}$
& $0.2_{-0.1}^{+0.3}$ & $1.7^{+0.5}_{-0.3} $& 0.85 & 1.2 \\
(vi)- {\it Swift} & $0.07^{+0.03}_{-0.04}$ & $7^{+0.8}_{-2}$
& $0.2_{-0.1}^{+0.3}$ & $2^{+0.2}_{-0.3} $& 0.9 & 1.2 \\
  \hline
\end{tabular}
\caption{ Best fit parameters $Rate(z=0)$ , $L^*$, $\alpha$ and
$\beta$ and their 1-$\sigma$ confidence levels. For each fit we
report the  $\chi^2$ values corresponding to the best fit
($\chi^2_{\rm b.f.}$). Also shown are the $\chi^2$ values for the
fit to the BATSE({\it Swift}) data obtained using the  parameters
that best fit the {\it Swift}(BATSE) sample ($\chi^2_{\rm
other}$). }
\end{table}

\section{The redshift distributions}

We turn now to the observed redshift distributions of BeppoSAX/HETE2
and {\it Swift}.  For BeppoSax/HETE2 we consider the observed
distribution of all the  with an available redshift: 32 bursts from
http://www.mpe.mpg.de/~jcg/grbgen.html (excluding GRB980425 with
$z=0.0085$). For {\it Swift} we consider all bursts with an
available redshift: 39 bursts  from the {\it Swift} home page.

The redshift sample is influenced by selection effects that are hard
to quantify. To rectify this problem we compare our models both with
the raw data and with   corrected samples that attempt  take these
effects into account.  The selection effects are most severe if the
redshift determination depends on the identification of emission
lines in the spectrum of the host galaxy. Since there are very few
emission lines in the range $1.3<z<2.5$ (Hogg \& Fruchter, 1999)
such redshifts may be missed. For BeppoSAX/HETE2 this is the main
mode of redshift determination. We follow, therfore,  Hogg \&
Fruchter (1999) and consider a modified distribution in which all
 BeppoSAX/HETE2  GRBs with an optical afterglow but without a
redshift determination are assigned  uniformly in this redshift
range $1.3<z<2.5$. Using the data in
http://www.mpe.mpg.de/~jcg/grbgen.html we have a total sample of 46
BeppoSAX/HETE2  GRBs: 32 with measured redshift and 14 with no
measured redshifts which we assign to this range in a uniform way.
For completeness we also check what happens if all bursts with no
redshift but with optical afterglow are nearby, that is they are
distributed uniformly between
 $(0<z<1.5)$, or distant, i.e.  distributed uniformly between $z=2.3$ and the
maximal  BeppoSAX/HETE2 redshift $z=4.3$.

The situation concerning {\it Swift} bursts is more complicated as
most {\it Swift} redshifts are obtained using absorption lines  in
the optical afterglow. The main selection effect in this case is the
weakness of the afterglow signal or the optical depth within the
host (Fiore et al. 2007). Both selection
effects work against high redshift bursts.  Lacking a clear model we
consider for {\it Swift} just the observed data set.

\begin{figure}[h]
\begin{tabular}{cc}
\includegraphics[height=6.8truecm,width=6.8truecm]{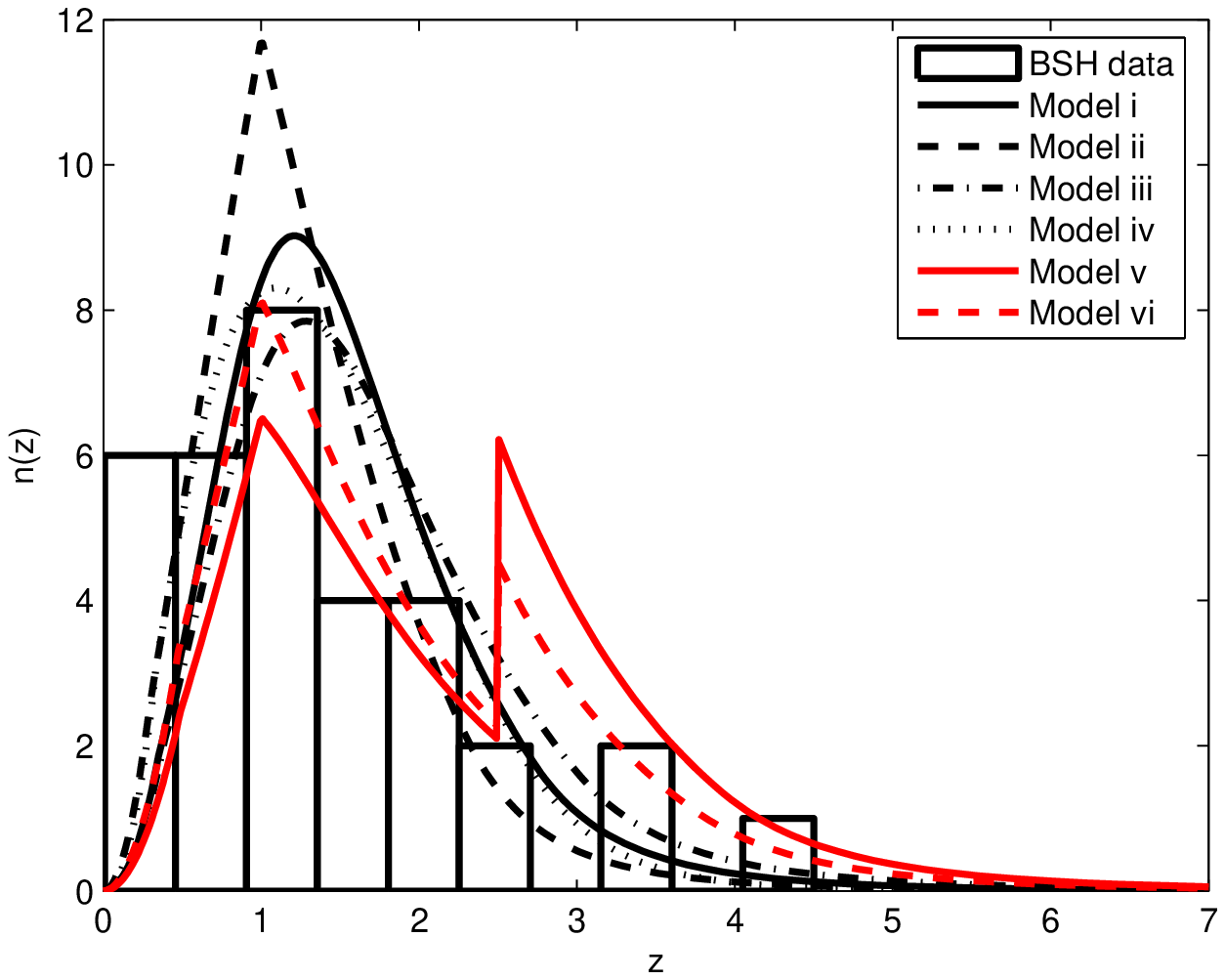}
\includegraphics[height=6.8truecm,width=6.8truecm]{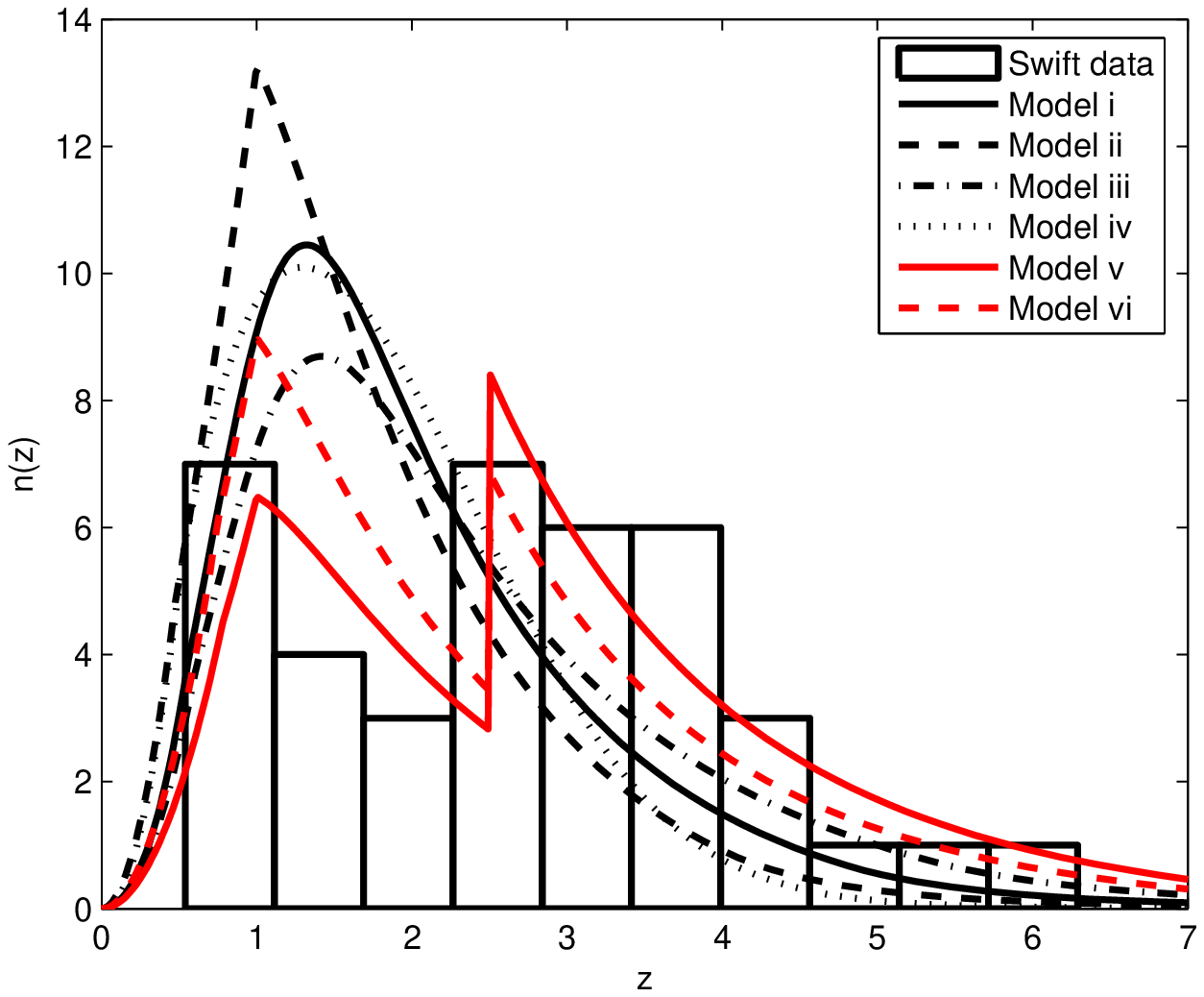}
\end{tabular}
\caption{The predicted and the observed differential distributions
of the GRBs redshift for the different models (a) left panel - {\it
Swift} data with theoretical models with  $P_{lim}=0.18$ (b) right
panel BeppoSax/HETE2 (BSH in the legend)  and the models with
$P_{lim}=0.5$.} \label{diff}
\end{figure}

Figure \ref{zdistBATSE} depicts the isotropic peak luminosities and
redshifts of the BeppoSAX/HETE2 and {\it Swift} samples. This figure
shows clearly the differences in thresholds.  In the second figure
we also plot the values of L for the average photon index assumed in
the calculations. As we can see from this figure the values of the
peak luminosities obtained using the average spectrum are rather
similar to the ones obtained using the real spectrum. Therefore, it
is reasonable  to use the average spectrum for the k-correction as
we have done in this analysis. Another feature seen in this figure
is that the {\it Swift} redshift distribution shows (seen even more
clearly in Fig. \ref{diff}) a paucity of bursts in the range $1 < z
< 2$. It is not clear if this is statistically significant, but it
is apparent in the data. There is no clear selection effect that
could give rise to this feature.

Using the different models for the GRB rate and the luminosity
function we derive now the expected distribution of the observed
bursts' redshifts:
\begin{equation} \label{red} n(z,L) dz
d\log(L)  = \frac{R_{GRB}(z)}{1+z} \frac{dV(z)}{dz} \Phi_o(L)d\log
L \ \ .
\end{equation}
The expected redshift distribution is:
\begin{equation}
\label{redshift} N(z)= \frac{R_{GRB}(z)}{1+z} \frac{dV(z)}{dz}
\int_{L_{\rm min}(P_{\rm lim},z)}^{L_{\rm max}} \Phi_o(L)d\log L \
 ,
\end{equation}
where $L_{\rm min}(P_{\rm lim})$ is the luminosity corresponding to
minimum peak flux $P_{\rm lim}$ for a burst at redshift z and
$L_{\rm max}=L^*\times \Delta_2=10\,L^*$. This  minimal peak flux
corresponds to the detector's threshold. For BeppoSax/HETE2 we use,
$P^{(50-300) keV}_{\rm lim}\sim 0.5 $ ph cm$^{-2} s^{-1}$, which is
roughly the limiting flux for the GRBM on BeppoSAX (Guidorzi 2002).
 The triggering algorithm for {\it Swift} is rather complicated but as
shown in Fig. 1 it can be approximated by a minimal rate:
$P^{(50-300) keV}_{\rm lim Swift} \sim 0.18$ ph/cm$^{2}$/sec.

\begin{figure}[h]
\begin{tabular}{cc}
\includegraphics[height=6.8truecm,width=6.8truecm]{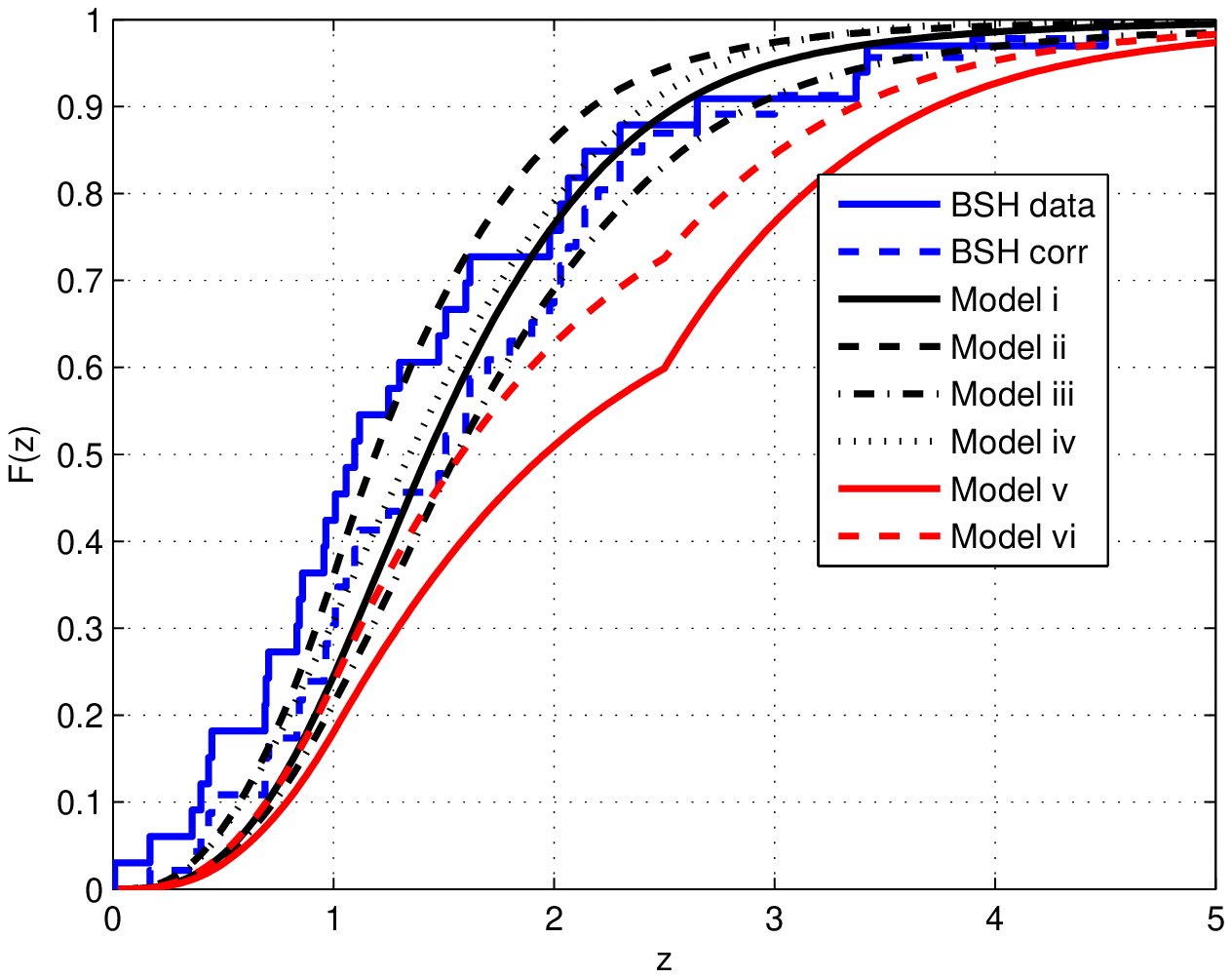}
\includegraphics[height=6.8truecm,width=6.8truecm]{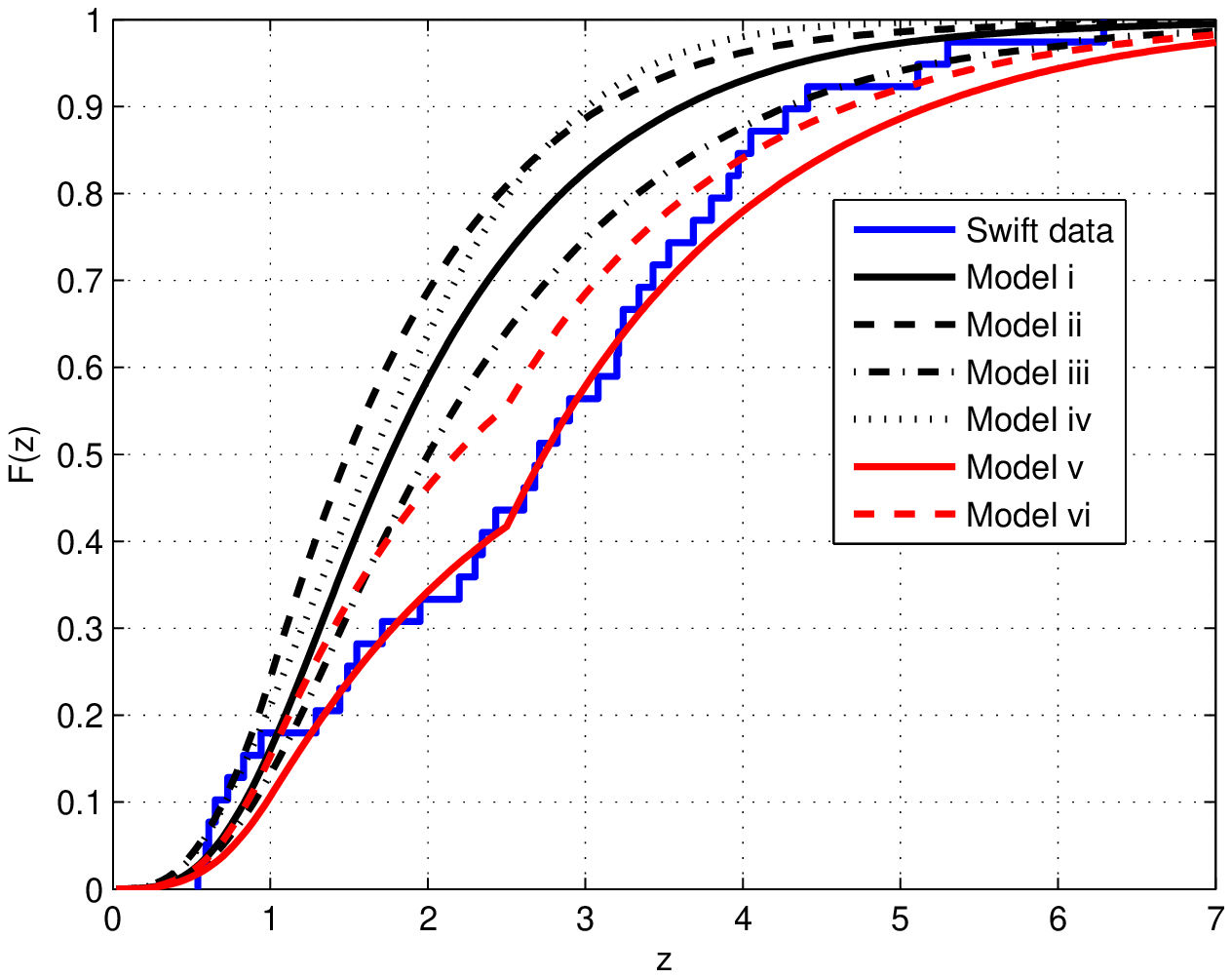}
\end{tabular}
\caption{ The predicted and the observed cumulative distributions
of the GRBs redshift for the different models (a) left panel -
{\it Swift} data with theoretical models with  $P_{lim}=0.18$ (b)
right panel BeppoSAX/HETE2  and the models with $P_{lim}=0.5$. For
BeppoSAX/HETE2 we also show the  distribution where selection
effects are taken into account assuming that all the GRB with no
redshift but with optical afterglow lie in the range $1.3<z<2.5$.}
\label{integrated}
\end{figure}

Our results are summarized in Table 2  and in Figs. \ref{diff} and
\ref{integrated} which  present a comparison of the observed
(corrected and uncorrected) BeppoSAX/Hete2 and {\it Swift} redshift
distributions with theoretical models that were obtained from best
fits of the model's parameters to the {\it Swift} peak flux
distribution. Qualitatively similar results are obtained from best
fits to the  BATSE peak flux distribution. Fig. \ref{diff} depicts
the differential distribution of the observed redshifts while Fig.
\ref{integrated} depicts the integrated distribution. The values of
the KS test for the different models are shown in Table 2.

The most remarkable feature is that none of the pure SFR models
(i-iv) is consistent with the {\it Swift} data. This result is
consistent with the findings of Daigne et al. (2006). The {\it
Swift} redshift distribution is inconsistent even with model (iii),
that is model SF3 of Porciani and Madau (2001) which is rising at
large redshifts. It is consistent only with distributions (v) and
(vi) that involve an ``artificial" large redshift enhancement
compared to the standard SFR model.

Consequently, it is difficult to find models that fit both the {\it
Swift} and the {BeppoSAX/HETE2} data. Models (i), (iii) and (iv)
that favor a nearer GRB distribution, are consistent with the
BeppoSAX/Hete2 distribution while model (v) that favors a more
distant distribution is consistent with the observed {\it Swift}
redshift distribution. The only combined fit to both data sets is
obtained for model (vi) which is consistent with the {\it Swift}
distribution (KS values $\sim$ 0.19) and with the corrected the
BeppoSAX/HETE2 data (KS values $\sim$ 0.17).  Model (vi) represents
a variation of the rather arbitrary parameters of model (v).
Clearly, we can consider a series of models based on RR SFR ranging
from  no  enhancement at high z (model ii) which fits the
uncorrected BeppoSAX/HETE2 sample to  a very strong enhancement at
high z (model v) that fits just the {\it Swift} data. In model (vi)
we consider an intermediate enhancement which is formally consistent
with both the {\it Swift} and the modified BeppoSAX/HETE2 data.
However, as we discuss later, even in these models and even in the
models like, (i), (ii), (iv) that are compatible with BeppoSAX/HETE2
redshift distribution, the two dimensional redshift luminosity
distribution shows too many high luminosity bursts. Note that
similar results were obtained when we modified SF2 by adding an ad
hoc enhancement at large redshift.

\begin{figure}[h]
\includegraphics[height=6.8truecm,width=6.8truecm]{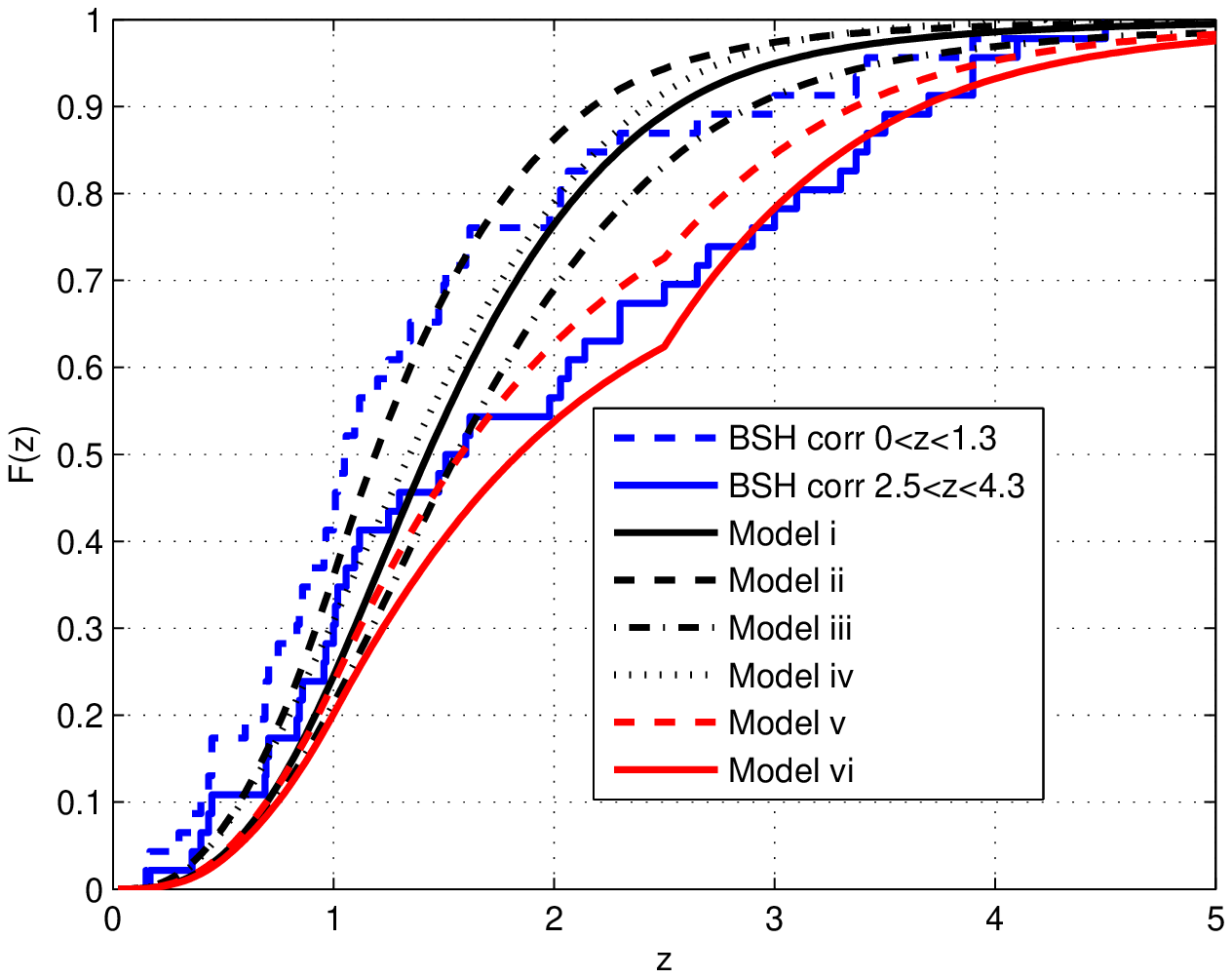}
\caption{ The predicted and the observed cumulative distributions
of the GRBs redshift for  BeppoSAX/HETE2 sample
 and the models with $P_{lim}=0.5$. We
 also show the  distribution where selection
effects are taken into account assuming that all the GRB with no
redshift but with optical afterglow lie in the range $0<z<1.3$ and
$2.3<z<4.3$.}
\label{corr}
\end{figure}

Fig.\ref{corr} depicts a comparison of the theoretical models with
two extreme modifications of the BeppoSAX/HETE2 distributions. These
modifications attempt to estimate different extreme effects of the
selection effects for BeppoSAX/HETE2. In one case we put all bursts
with an optical afterglow and an unknown redshift at a large
redshift (a uniform distribution in the range in the $2.3<z<4.3$).
In the other case we put all these bursts uniformly in z at small
redshift ($z<1.3$). The KS values of the comparison with different
models are shown in Table 3. We see that putting all the bursts
uniformly distributed in the range $0<z<1.3$ has a little effect.
Models (like (ii) and (iv)) that are consistent with the uncorrected
distribution are consistent with the corrected one and others are
now. Putting all bursts with an unknown redshift at $2.3<z<4.3$ has
a strong effect. With this correction no SFR can fit the
BeppoSAX/HETE2 corrected data and we need a model like (v) or (vi)
with a high redshift enhancement. This extreme (high redshift)
correction makes the BeppoSAX/HETE2 data compatible with the {\it
Swift} data. Note, however, that there is no clear reason to choose
such a correction.

\begin{table}[h]
\begin{tabular}{|c|c|c|c|}
 \hline
model
   & BeppoSAX/HETE2 &  BeppoSAX/HETE2 &  {\it Swift}\\
 &  &   $1.3 < z < 2.5$ correction&  \\
  \hline
(i)  & $0.04$ & $0.69$ & $<0.01$  \\
(ii) & $0.63$ & $0.03$ & $<0.01$ \\
(iii)& $0.01$ & $0.27$ & $0.03$   \\
(iv) & $0.25$ & $0.41$ & $<0.01$   \\
(v) & $<0.01$&$ 0.05$ & $0.85$  \\
(vi) & $0.03$&$ 0.17$ & $0.19$  \\ \hline
\end{tabular}
\caption{KS probability values for the BeppoSax/HETE2
 and the {\it Swift} samples for theoretical models with parameters that best
 fit the Swift peak flux distribution}
\end{table}

\begin{table}[h]
\begin{tabular}{|c|c|c|}
 \hline
model &  BeppoSAX/HETE2   &
BeppoSAX/HETE2  \\
& $0<z<1.3$ correction &
$2.3<z<4.3$ correction\\
  \hline
(i)  & $0.01$ & $0.02$  \\
(ii) & $0.50$ & $<0.01$ \\
(iii)& $<0.01$ & $0.14$ \\
(iv) & $0.10$ & $<0.01$ \\
(v)  & $ <0.01$ & $0.56$ \\
(vi) & $<0.01$ &  $0.18$ \\ \hline
\end{tabular}
\caption{KS probability values for two extreme corrections to the
BeppoSax/HETE2 sample. The low redshift correction adds all the
bursts with optical afterglow and unknown redshift uniformly in the
region $0<z<1.3$ and the high redshift correction adds them
uniformly at the region  $2.3<z<4.3$. The theoretical models use
parameters that best fit the {\it Swift} peak flux distribution}
\end{table}

\begin{figure}[h]

\includegraphics[height=6.8truecm,width=6.8truecm]{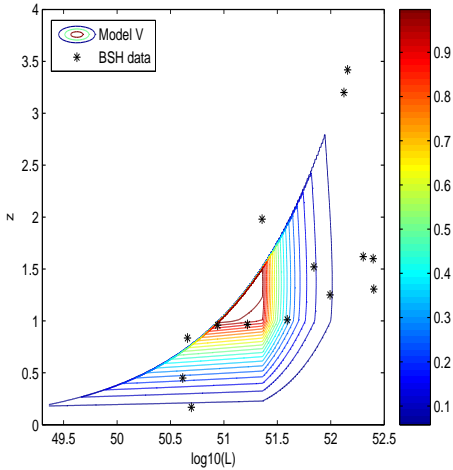}
\caption{ The two dimensional probability distribution of expected
redshift and luminosity for the luminosity function parameters that
best fit the {\it Swift} peak flux distribution considering a RR-sfr
and $P_{lim}=0.5$ph/cm$^2$/s. Contour lines are 0.9,0.8... 0.01  of
the maximum. Also marked are the BeppoSax/HETE2 (BSH in the legend)
GRBs with a known redshift and spectral index. Note that there is
only one burst in the ''forbidden" region in the upper left part of
the plot where the peak flux is below $0.5$ph/cm$^2$/s.}
\label{2dBHS}
\end{figure}

Fig. \ref{2dBHS} depicts a comparison of the two dimensional
redshift and luminosity distributions between the BeppoSAX/HETE2
data and model (ii). Naturally, we include here only bursts with
known redshifts. Several features are apparent. First, the estimate
of $P_{lim} =0.5$ for BeppoSAX/HETE2 is reasonable. Only one burst
is detected in the ``forbidden" region with a lower peak flux.
However, it is clear that there is no good fit between the model and
the observed distribution. The lack of bursts in the range $1.5 < z
< 3 $  may be explained by selection effects. However, in addition,
there are significantly more high luminosity bursts than predicted
by the model. It is clear that even though the KS test of the
integrated redshift distribution for this model suggests that the
model is consistent with the observed distribution the two
dimensional distribution of luminosities and redshifts is
inconsistent. Similar, or worse results are obtained for this data
with all other models  that we have considered including, in
particular, model (vi).

\begin{figure}[h]
\includegraphics[height=6.8truecm,width=6.8truecm]{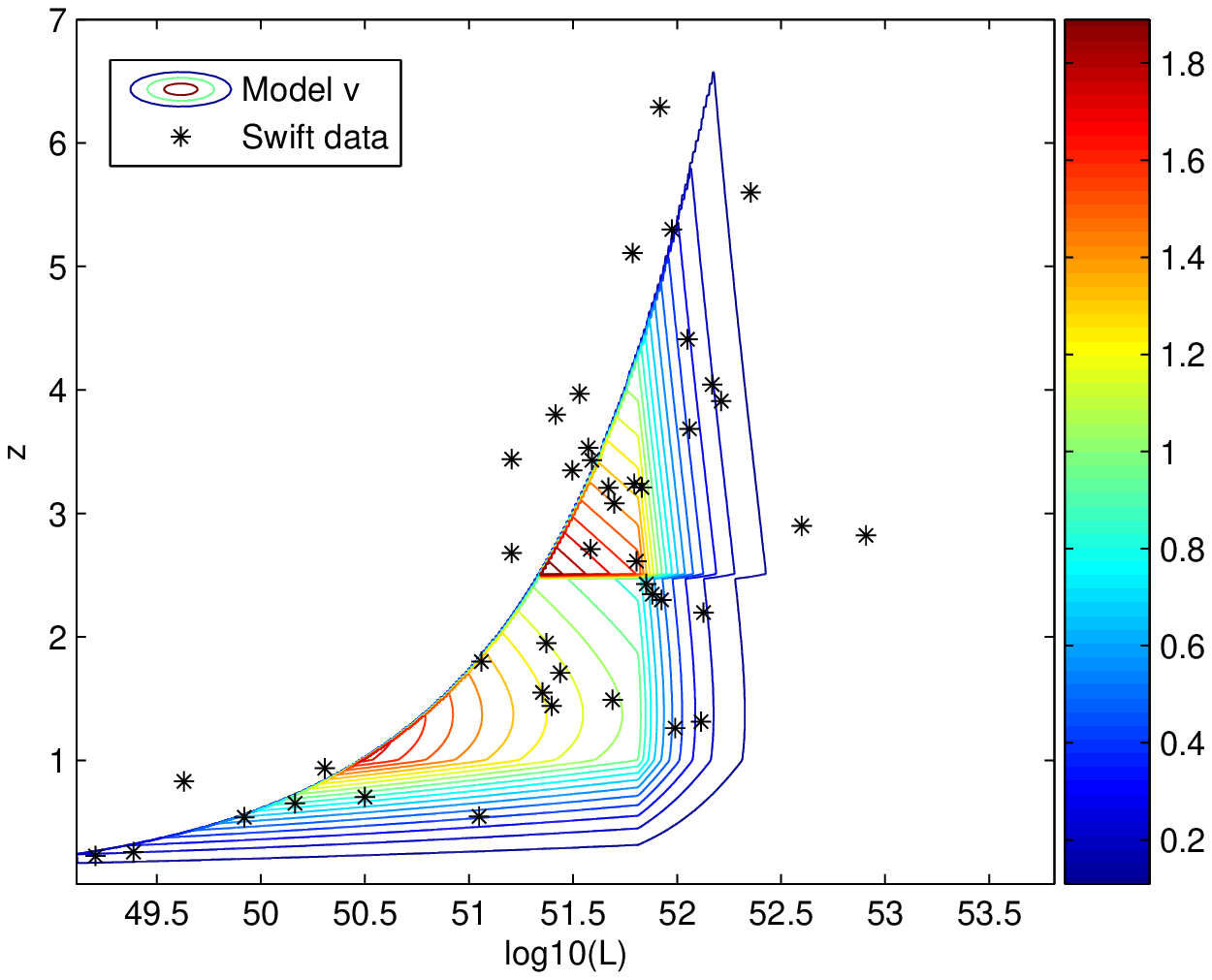}
\caption{  The two dimensional probability distribution of expected
redshift and luminosity for the luminosity function parameters that
best fit the {\it Swift} peak flux distribution considering model
(v) and $P_{lim}=0.18$ph/cm$^2$/s.   Contour lines are 0.9,0.8...
0.01 of the maximum. Also marked are the {\it Swift} GRBs with a
known redshift. Note that there are several  bursts in the
''forbidden" region in the upper left part of the plot where the
peak flux is below $0.18$ph/cm$^2$/s. These bursts reflect the fact
that {\it Swift}'s trigger is not based just on peak flux counts.
However as these bursts cluster very close to the line
$P_{lim}=0.18$ph/cm$^2$/s we find that the complicated triggering
algorithm of {\it Swift} if not an issue here.}
 \label{2dSwift}
\end{figure}

A similar comparison between model (v) and the {\it Swift} data is
shown in Fig \ref{2dSwift}. Here there are several bursts in the
forbidden region in the upper left part of the plot where the peak
flux is below $0.18$ph/cm$^2$/s. These bursts reflect the fact that
{\it Swift}'s trigger is not based just on peak flux counts. However
as these bursts cluster very close to the line
$P_{lim}=0.18$ph/cm$^2$/s we find that the complicated triggering
algorithm of {\it Swift} is not an issue. The fit of the observed
data to the model is clearly better than the one seen in Fig.
\ref{2dBHS}. Still it is not compelling. Here the basic problem can
be seen also in Fig. \ref{diff} that depicts the observed
differential redshift distribution of {\it Swift} bursts. The
paucity of bursts in the range $1 < z < 2$ hints towards a two
population model - or towards a high redshift enhancement of the
sort that we have crudely modeled in (v). Unlike the BeppoSAX/HETE2
sample we don't see here a significant fraction of high luminousity
bursts (as compared with the model) but again there are hints
towards a broader luminosity function that the one we use.


\section{Conclusions and Implications}

We find that {\it Swift} GRBs do not follow the SFR as described
by several different models (i-iv). Given these SFRs there is no
luminosity function that can fit both the {\it Swift} observed
peak flux and the z-distributions. We were able to obtain a
reasonable fit when we considered a GRB rate function  that
included a high redshift enhancement. This might be related to
suggestions  that long GRBs arise preferably in low metallicity
regions (Fynbo et al 2003; Vreeswijk et al., 2004). But it could
arise from other reasons. We used a simple toy scheme to model
this enhancement. Because of the very crude nature of the model
and the limited scope of the available data we did not try to
optimize extensively the parameters of this model. It was
reassuring that we obtained a reasonable fit with such a simple
model and without an extensive search for the parameters. It is
remarkable that the enhancement arises in the high redshift range,
where the bursts and their afterglow are weaker and hence
selection effects are expected to reduce rather than increase the
number of bursts with detected redshifts.

We mention now the strange  paucity of {\it Swift} bursts with $1 <
z < 2$.  If real  and not just a statistical fluctuation or a result
of an unknown selection effect this paucity may indicate: (a) A jump
in long GRB rate or another factor at higher redshifts; (b) A
dependence of the luminosity function on the rate or even the
existence; (c) The appearance of two populations one at lower
redshift and another one at higher redshifts (which can be viewed as
a special case of a z dependence of the luminosity function). While
these speculations are intriguing it is clear that it   is essential
to determine the selection effects that control the samples of GRBs
with determined redshifts before far reaching conclusions are made.

The only GRB rate and luminosity functions that are consistent with
these distributions and with both observed redshift distributions
(of BeppoSAX/HETE2 and of {\it Swift}) is a one with an enhanced GRB
rate at large redshift. This consistency is achieved only  after we
modified the BeppoSAX/HETE2 sample by adding all bursts with no
redshift in the range $1.5 < z < 2.5$ in which there are no strong
emission lines and redshift identification is difficult (Hogg \&
Fruchter, 1999) or if we put all those bursts, artificially, at high
redshifts $2.3 < z < 4.5 $. However, as the two dimensional
distribution of redshifts and luminosities of the Bepposax/HETE2
does not seem to fit the model we do not assign a great significance
to this fact.

Another important result is that the BATSE and {\it Swift} peak flux
distributions are consistent with each other and with the estimated
limiting fluxes for detection for the two detectors. The combined
analysis suggests that the local rate of GRBs (without a beaming
correction) can be determined up to a factor of approximately five
and it ranges between 0.05Gpc$^{-3}$yr$^{-1}$ for a rate function
that has a large fraction of high redshift bursts to
0.27Gpc$^{-3}$yr$^{-1}$. Note that the inferred low local rates,
$\sim 0.05$ Gpc$^{-3}$yr$^{-1}$, which are about an order of
magnitude lower than previous estimates (Guetta, Piran \& Waxman,
2005), arise from the models that involve metallicity enhancement at
large redshifts. These rates do not include the beaming correction
which is of order $\sim 100$. Even with this correction these rates
correspond to a local rate of a burst per $2 \times 10^6$ years per
galaxy, indicating that strong GRBs are a very rate phenomenon.
However, the actual rate of weak bursts could be much higher if
indeed there is a large population of very low luminosity bursts, as
inferred from the detection of GRB 060218 (Soderberg et al., 2006).
These models predict, on the other hand, a relatively large fraction
of about 6\% of high redshift $(z>6)$ {\it Swift} bursts.

When considering the BeppoSAX/HETE2 redshift distribution on its own
it seemed that even luminosity functions and rates that fit the peak
flux distribution and the observed redshift distributions do not fit
the two dimensional luminosity and redshift distribution. The
paucity of bursts with redshifts between $1.5 < z< 2.5$ can be
explained by the selection effect mentioned earlier. However, the
excess of very luminous low redshift bursts is unexpected and
indicates that other selection effects that favor identification
redshift determination of such bursts take place and possibly
dominate the BeppoSAX/HETE2 sample. If correct this has potential
implications to other statistical information that has been
determined from this data sample.

\end{document}